\documentclass[printer]{aa} 
\bibpunct{(}{)}{;}{a}{}{,} 
\usepackage{graphicx}
\usepackage{txfonts}
\begin{document}

   \title{Temporal Evolution of the Size and Temperature of Betelgeuse's Extended Atmosphere}

   \author{E. O'Gorman
          \inst{1}\fnmsep\thanks{eamon.ogorman@chalmers.se},
          G. M. Harper
          \inst{2},
          A. Brown
          \inst{2},
          E. F. Guinan
          \inst{3},
          A. M. S. Richards
          \inst{4},
          W. Vlemmings
          \inst{1},
          \and
          R. Wasatonic
          \inst{3}
			}
   \institute{Department of Earth and Space Sciences, Chalmers University of Technology, Onsala Space Observatory, 439 92 Onsala, Sweden
          \and
             Center for Astrophysics and Space Astronomy, University of Colorado, 389 UCB, Boulder, CO 80309, USA 
		 \and
		 	 Department of Astrophysics and Planetary Science, Villanova University, Villanova, PA 19085, USA
	 	 \and
		     Jodrell Bank Centre for Astrophysics, School of Physics and Astronomy, University of Manchester, Manchester M13 9PL, UK     
             }
 
  \abstract{Spatially resolved multi-wavelength centimeter continuum observations of cool evolved stars can not only constrain the morphology of the radio emitting regions, but can also directly probe the mean gas temperature at various depths of the star's extended atmosphere. Here, we use the Very Large Array (VLA) in the A configuration with the Pie Town (PT) Very Long Baseline Array (VLBA) antenna to spatially resolve the extended atmosphere of Betelgeuse over multiple epochs at 0.7, 1.3, 2.0, 3.5, and 6.1\,cm. The extended atmosphere deviates from circular symmetry at all wavelengths while at some epochs we find possible evidence for small pockets of gas significantly cooler than the mean global temperature. We find no evidence for the recently reported e-MERLIN radio hotspots in any of our multi-epoch VLA/PT data, despite having sufficient spatial resolution and sensitivity at short wavelengths, and conclude that these radio hotspots are most likely interferometric artefacts. The mean gas temperature of the extended atmosphere has a typical value of 3000\,K at 2\,$R_{\star}$ and decreases to  1800\,K at 6\,$R_{\star}$, in broad agreement with the findings of the single epoch study from \cite{lim_1998}. The overall temperature profile of the extended atmosphere between $2\,R_{\star} \lesssim \mathit{r} \lesssim 6\,R_{\star}$ can be described by a power law of the form $T_{\mathrm{gas}}(r)\propto r^{-0.6}$, with temporal variability of a few 100\,K evident at some epochs. Finally, we present over 12 years of V band photometry, part of which overlaps our multi-epoch radio data. We find a correlation between the fractional flux density variability at V band with most radio wavelengths. This correlation is likely due to shock waves induced by stellar pulsations, which heat the inner atmosphere and ionize the more extended atmosphere through radiative means. Stellar pulsations may play an important role in exciting Betelgeuse's extended atmosphere.}
   \titlerunning{The Temporal Evolution of Betelgeuse's Extended Atmosphere.}
   \authorrunning{E. O'Gorman et al.}
   \maketitle
\section{Introduction}
\label{sec1}
Our understanding of the mass-loss mechanism in early-M spectral type red supergiants (RSGs) has been inhibited by the lack of spatially resolved observations on the scale of the chromosphere and wind acceleration region; the two regions identified as the most important for studies of mass-loss mechanisms in evolved stars \citep{holzer_1985}. Although radiation pressure on dust may play a role in driving the wind to its terminal velocity via dust-gas collisions, additional mechanisms are required to accelerate the gas out to distances where substantial dust condensation may occur \citep[e.g., ][]{harper_2009}. Thermal free-free centimeter continuum emission arises from the star's extended atmosphere and directly probes the crucial inner spatial scales containing the chromosphere and the initial wind acceleration region. Spatially resolved centimeter observations not only allow a study of the gas morphology but also enable the mean gas temperature to be empirically derived as a function of depth through the atmosphere \citep{lim_1998} (hereafter L+98). It is therefore a powerful tool to study the details of mass loss in RSGs. Despite the value of spatially resolved centimeter continuum observations, very few RSGs have sufficiently large angular diameters that they can be resolved with current centimeter interferometers.

Betelgeuse ($\alpha$ Ori, M2\,Iab) is the closest isolated RSG \citep[$d=197 \pm 45$\,pc;][]{harper_2008} and is therefore the best studied of the early-M RSG class. Its has the largest continuum angular diameter of all RSGs ($\phi _{\star} = 42.49 \pm 0.06\,$mas at $2.3\,\mu$m, \citealt{ohnaka_2011}) and its extended atmosphere has been the focus of many centimeter studies over the past few decades. The first detailed study of Betelgeuse at centimeter wavelengths was carried out by \cite{newell_1982} with the Very Large Array (VLA) in its second most compact C configuration, at a resolution worse than $\sim 1\arcsec$. The atmosphere was unresolved but the radio emission was interpreted as chromospheric in origin and extending from 1 to 4\,R$_{\star}$. This was in broad agreement with subsequent Alfv\`en wave models \citep{hartmann_1984} and later Hubble Space Telescope (\textit{HST}) spatially resolved ultraviolet observations \citep{gilliland_1996,uitenbroek_1998}. Spatially resolved VLA plus Multi-Element Radio Linked Interferometer Network (MERLIN) observations at 6\,cm also confirmed the extended nature of the radio emitting region \citep{skinner_1997}. 

L+98 used the VLA in its most extended A configuration to resolve Betelgeuse's atmosphere at 0.7\,cm (with 40\,mas resolution) and partially resolve it at 1.3, 2.0, 3.5, and 6\,cm (with resolution between 90 and 330\,mas). Because the radio emission is thermal and optically thick, they were able to derive the mean gas temperature as a function of radius, and discovered that the temperature never reached chromospheric values (i.e., $\sim 4000-8000$\,K) but decreased steadily from $\sim$3450\,K at 2\,R$_{\star}$, to $\sim$1370\,K at 7\,R$_{\star}$. They also detected an asymmetry in their 0.7\,cm image which they attributed to the action of a large convection cell. To reconcile their results with the extended ultraviolet observations, which probed the same regions, they concluded that the inner atmosphere must be inhomogeneous to accommodate the hot chromospheric plasma, but that the cooler gas must be 3 orders of magnitude more abundant. \cite{harper_2006} used observations of the chromospheric electron density tracer C II] $\lambda 2325\,\AA$ to confirm this low filling factor for the chromospheric gas. 

Recently, an unexpected discovery by \cite{richards_2013} (hereafter R+13) with e-MERLIN at $80$\,mas\,$\times\,60\,$mas resolution, revealed that two unresolved radio features contribute about 75\% of the total flux density at 5.2\,cm. The two ``radio hotspots'' are separated by $90\pm 10$\,mas (i.e., $\sim 4\,$R$_{\star}$) and have inferred brightness temperatures $T_{b} = 3800\pm 500\,K$ and $T_{b} = 5400\pm 600\,K$, with the later value significantly above the photospheric temperature ($T_{\rm{eff}}=3690\pm 54\,$K, \citealt{ohnaka_2011}). Indeed, because the features are unresolved, and assuming the emission is thermal and optically thick, then these inferred brightness temperatures will be lower limits for the gas temperature. The potential importance of these radio hotspots for understanding the wind acceleration region of RSGs has been the motivation for this paper. In the following sections we present multi-epoch multi-wavelength radio observations of Betelgeuse, taken $\sim 10$ years prior to these e-MERLIN observations. At short wavelengths, our data have comparable or superior spatial resolution to the e-MERLIN data and so the goal of this paper is to search for signatures of hotspot features to improve our understanding of Betelgeuse's wind acceleration region.

\section{Observations and data reduction} 
Betelgeuse was observed with the NRAO\footnote{The National Radio Astronomy Observatory is a facility of the National Science Foundation operated under cooperative agreement by Associated Universities, Inc.} VLA  in the A configuration with the Pie Town (PT) Very Long Baseline Array (VLBA) antenna at multiple epochs between December 2000 and October 2004. The PT antenna is located approximately 52\,km from the center of the VLA, and its inclusion to the A configuration enhances the east-west resolution by a factor of 2 \citep{ulvestad_1998}. The dates and corresponding combinations of wavelengths used in our observations are summarized in the first two columns of Table \ref{tab1}. Our two multi-wavelength data sets in 2002 (Program: AH0778) were taken only $\sim 2$ months apart and consist of all wavelength bands between 1.3 and 20.5\,cm. We observed the star again under program AH0824 in 2003 and 2004 at all wavelength bands between 0.7 and 20.5\,cm. We also obtained VLA A configuration plus PT antenna data from the NRAO data archive (Program: AL0525) at  0.7 and 1.3\,cm which were taken in late 2000 and early 2001, respectively.

We used the \textit{fast switching} phase calibration technique at 0.7, 1.3, and 2.0\,cm to compensate for tropospheric phase variations \citep{carilli_1996}, switching between observations of the point source phase calibrator 0532+075 (located 5.6$^{\circ}$ from Betelgeuse) and Betelgeuse, with cycle times varying between 2 and 3\,min. The archival data from 2000/2001 also implemented the fast switching technique at these wavelengths but used the point source phase calibrator 0552+032, which was $\sim 10$ times fainter at 0.7\,cm but 4.2$^{\circ}$ from Betelgeuse. We also interleaved observations of 0532+075 to calibrate phase variations at longer wavelengths, but implemented longer cycle times of $10-20$\,min. Absolute flux density calibration was obtained from 3C\,48 at 6.1\,cm, and 3C\,286 at all other wavelengths. The absolute flux density uncertainty at all epochs is estimated to range from 3\% to 10\% between the longest to shortest wavelengths, respectively. The flux density uncertainties quoted in the remainder of this paper do not include this systematic error but represent the $\pm 1\sigma _{\mathrm{rms}}$ random error. The raw visibilities were flagged and calibrated according to standard VLA continuum data reduction procedures using the Astronomical Image Processing System \cite[AIPS;][]{greisen_1990}. The calibrated visibilities were imaged in both AIPS and the Common Astronomical Software Application  \cite[CASA;][]{mcmullin_2007} package.

\subsection{Imaging}\label{obs2.1}
The calibrated visibilities at all wavelengths were imaged using CASA's \textit{clean} task while applying uniform weighting. Additional images were also created at 0.7 and 1.3\,cm which implemented both Briggs weighting \citep{briggs_1995}, and natural weighting with a restoring beam size equal to that obtained using uniform weighting, as used in L+98. These additional images were used to check if any asymmetries produced by one weighting scheme could be reproduced by another.

\subsection{Calibrated visibilities}\label{obs2.2}
Betelgeuse was observed offset by a few restoring beams from the phase reference center to avoid any possible spurious artefacts which might accumulate there. One such artefact was noticed in the 2.0\,cm image from April 2002 but our target was sufficiently far away and was not contaminated. We fitted uniform elliptical Gaussian brightness distributions to the source in each image and measured the position of the peak emission. We then used CASA's \textit{fixvis} task to shift the visibilities such that this peak emission position was placed at the phase reference center. This shift enabled us to extract the azimuthally averaged complex visibilities discussed later in Section \ref{sec3.2}. The Python-based task \textit{uvmultifit} \citep{marti_vidal_2014} was then used to fit uniform intensity disks to the calibrated visibilities, with the major axis ($\theta _{\rm{maj}}$), axis ratio ($\theta _{\rm{min}}/\theta _{\rm{maj}}$, where $\theta _{\rm{min}}$ is the minor axis), position angle (PA), and flux density ($F_{\nu}$) left as free parameters. 

\begin{table*}
\caption{Multi-epoch VLA A configuration plus PT antenna observations of Betelgeuse.}
\label{tab1}
\centering
\begin{tabular}{c c c c c c c c c}
\hline\hline
Date						& $\lambda$			    & Restoring beam            & $\sigma_{\mathrm{rms}}$      		& $\theta _{\mathrm{maj}}$  &  $\theta _{\mathrm{min}}$	& PA  &
$F_{\nu}$ & $T_{b}$ \\
     			& (cm)                        		&  (mas $\times$ mas, $^{\circ}$)   	& (mJy\,beam$^{-1}$)                    	& (mas)   		    & (mas)   		    & ($^{\circ}$)   		    & (mJy)   		    & (K)	\\
\hline
\rule{-2.6pt}{2.5ex} 2004 Oct 21,30  & 0.7 & $39\times 26, 42$ & 0.37	& $99\pm 3$ & $92 \pm 3$& $92\pm 20$ &$28.68\pm 0.53$ & $2940\pm 140$\\
		 				 & 1.3 & $80\times 42, 32$ & 0.09   & $127\pm 2$& $ 114\pm 2$ & $85\pm 5$& $13.83\pm 0.10$& $3160\pm 80$\\
						 & 2.0 & $121\times 91, 54$ & 0.08  & $158\pm 6$& $ 152\pm 10$ & $132\pm 59$ & $7.23\pm 0.15$& $2360\pm 190$ \\
						 & 3.5 & $208\times 126, 41$ & 0.02	& $218\pm 6$& $182\pm 9$ & $162\pm 7$& $3.34\pm 0.03$& $2140\pm 120$\\
						 & 6.1 & $377\times 264, 46$ & 0.02	& $315\pm 29$& $187\pm 40$ & $173\pm 9$ & $1.55\pm 0.04$& $1920\pm 450$\\
						 & 20.5& $1262\times 889, 45$ & 0.03& $< 889$ & $< 889$ & $\dots$ &$0.25\pm 0.03$ &$ \textgreater 260$\\
\hline
\rule{-2.6pt}{2.5ex}  2003 Aug 10,12 & 0.7 		& $40\times 27, 47$ & 0.46	& $103\pm 4$& $92\pm 5$ & $104\pm 16$& $28.05\pm 0.84$& $2760\pm 200$ \\
									 & 1.3		& $80\times 42, 32$ & 0.17	& $122\pm 5$& $114\pm 6$& $8\pm 24$& $11.20\pm 0.24$& $2670\pm 190$\\
									 & 2.0		& $119\times 96, 62$& 0.10	& $132\pm 10$& $ 116\pm 9$& $11\pm 27$& $5.88\pm 0.17$&$3010\pm 340$\\
									 & 3.5		& $ 204\times 139, 46$&	0.03& $193\pm 7$& $ 140\pm 11$ & $152\pm 7$& $2.80\pm 0.04$&  $2630\pm 230$\\
									 & 6.1 		& $378\times 297, 69$&	0.03& $247\pm 47$& $179\pm 61$ & $169\pm 28$& $1.22\pm 0.04$&$2010\pm 780$ \\
					                 &20.5		& $1247\times 931, 49$&0.04	& $< 931$& $< 931$ & $\dots$& $0.26\pm 0.03$& $ \textgreater 250$\\
\hline
\rule{-2.6pt}{2.5ex}  2002 Apr 12,13  & 1.3 		& $91\times 59, 45$ & 0.18	&$134\pm 9$ & $102\pm 6$& $36\pm 10$& $8.96\pm 0.24$& $2170\pm 200$\\
							& 2.0		&$131\times 98, 60$ & 0.39	& $166\pm 16$& $104\pm 13$&$41\pm 11$ & $5.32\pm 0.24$& $2420\pm 400$ \\
							& 3.5		& $224\times 155, 65$&0.03	& $234\pm 9$& $171\pm 10$& $41\pm7$& $2.66\pm 0.04$& $1690\pm 120$\\
							& 20.5	& $1398\times 1146, 55$& 0.06	& $< 1146$ & $< 1146$ & $\dots$& $0.38\pm 0.06$& $ \textgreater 240$\\
\hline
\rule{-2.6pt}{2.5ex}  2002 Feb 17,18 & 1.3 		& $83\times 48, 30$&0.14	&$120\pm 4$ & $109 \pm 3$& $31\pm 13$&$10.87\pm 0.17$ & $2750\pm 130$\\
									& 2.0		& $128\times 90, 51$&0.11	&$148\pm 11$& $125\pm 18$& $142\pm 26$&$5.38\pm 0.22$ & $2280\pm 380$\\
									& 3.5		& $200\times 135, 43$&	0.03&$222\pm 8$ & $166\pm 13 $& $ 155\pm 7$ & $2.85\pm 0.04$& $1960\pm 170$\\
									& 20.5		&$1312\times 951, 55$ &0.05	& $< 951$& $< 951$& $\dots$& $0.30\pm 0.05$& $ \textgreater 270$\\
\hline
\rule{-2.6pt}{2.5ex}  2001 Jan 02  & 1.3 		& $78\times 42, 35$&0.08	&$124\pm 2$ & $115\pm 1$& $40\pm 8$& $12.58\pm0.08$& $2920\pm 60$ \\
 		 2000 Dec 23 & 0.7		& $44\times 20, 29$& 0.18	&$98\pm 2$ &$91\pm 1$ &$178\pm 8$ & $29.02\pm 0.30$& $3040\pm 80$\\
\hline
\end{tabular}
\tablefoot{The restoring beam properties (major axis, minor axis, PA) and image rms noise values are taken from the uniformly weighted radio images that include the PT antenna baselines. The major and minor axis of the resolved stellar radio disk $\theta _{\mathrm{maj}}$ and $\theta _{\mathrm{min}}$, the position angle PA (measured east of north), and the total flux density $F_{\nu}$, are all derived from the best-fit uniform intensity disk models using \textit{uvmultifit}. The atmosphere at 20.5\,cm was unresolved so we define the minor axis of the restoring beam as the upper limit to its diameter. The brightness temperatures, $T_b$, have been derived using Equation \ref{eq1} in Section \ref{sec3.4}.}
\end{table*}

\section{Results}\label{sec3}
\subsection{Radio images}\label{sec3.1}
Radio images were produced at all wavelengths using uniform weighting to utilize the full capabilities of the PT baselines and achieve maximum spatial resolution. The corresponding restoring beam sizes and image root mean square (rms) noise levels are provided in columns 3 and 4 of Table \ref{tab1}. The images were generally of high quality with the source clearly resolved at the shorter wavelengths. There were no obvious large scale asymmetries similar to those detected by R+13 present in any of the images at any wavelength. To investigate the presence of possible smaller scale asymmetries, analysis was carried out on the visibilities as the source to beam size ratio was small and this is discussed in Section \ref{sec3.2}. The 0.7\,cm images were the only ones to contain any obvious small scale (i.e., $\lesssim 40\,$mas) asymmetries. However, when these data sets were imaged using the different image weighting schemes specified in Section \ref{obs2.1}, the morphology was inconsistent. For example, the images created using standard uniform weighing were morphologically different to the images created using natural weighting with a restoring beam of size corresponding to that obtained with uniform weighting. These inconsistencies were probably due to poor phase stability resulting from rapid tropospheric phase fluctuations which our cycle times were not short enough to compensate for. To investigate this further, we removed the PT antenna baselines and again imaged the data using the two aforementioned imaging techniques. We found that two 0.7\,cm data sets (i.e., from 2000 and 2004) produced images that had consistent morphology when any of the imaging techniques were applied and showed no asymmetries. The observing logs of the other data sets report poorer weather conditions in agreement with the conclusions from our imaging analysis. 

\begin{figure*}
\includegraphics[trim = 6mm 0mm 0mm 0mm, clip,scale=0.95]{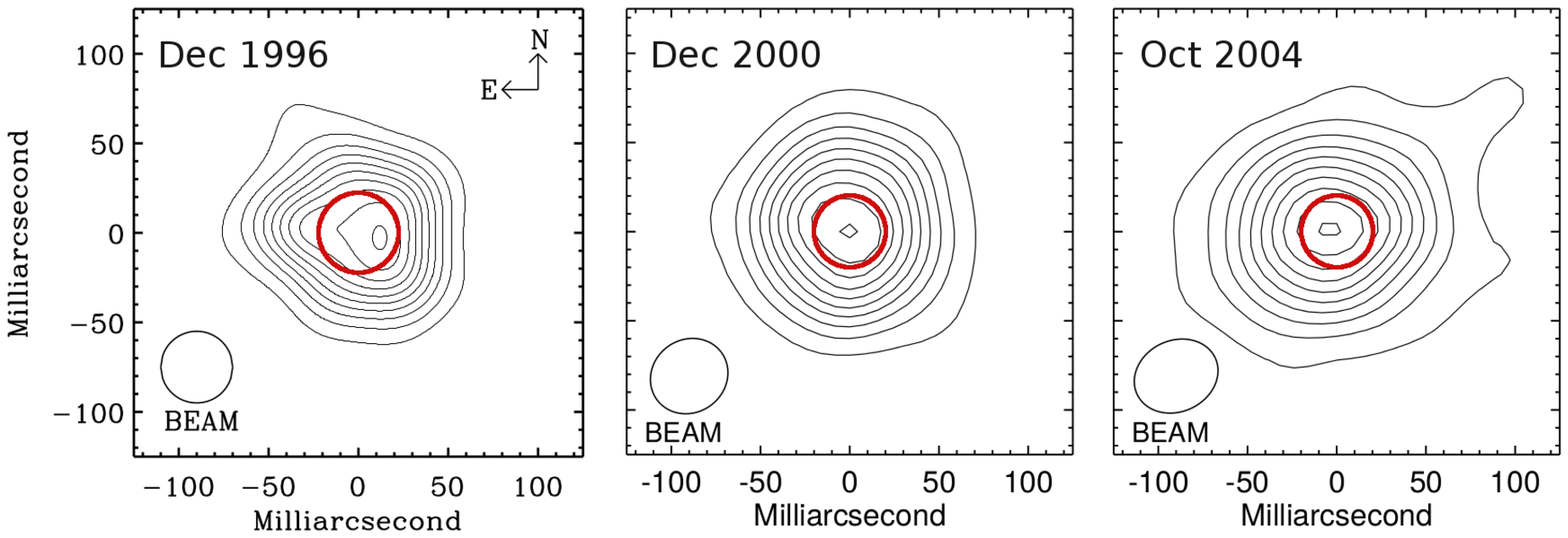}
\caption{VLA A configuration images of Betelgeuse at 0.7\,cm at three different epochs. These images have been created by naturally weighting the visibilities and applying a restoring beam corresponding to the size obtained from uniform weighting. The restoring beam is located in the bottom left corner of each panel while the red circle is the approximate location and size of the optical photosphere. The contour levels are plotted at $10\%, 20\%,\dots ,90\%$, and $99\%$ of the peak flux density. From left to right the beam sizes are $40^{\prime\prime}\times 40^{\prime\prime}$, $45^{\prime\prime}\times 41^{\prime\prime}$, and $48^{\prime\prime}\times 40^{\prime\prime}$. \textit{Left panel}: The L+98 image shows an asymmetry to the east of the photosphere which is here assumed to be coincident with the intensity-weighted centre of the radio disk. A photospheric angular diameter of 45\,mas \citep{dyck_1996} was assumed by the authors. \textit{Middle and right panels}: Our images show no large scale asymmetries. The position of the photosphere is here assumed to be located at the peak of the emission and we adopt the photospheric angular diameter of 42.49\,mas \citep{ohnaka_2011}}.
\label{fig1}
\end{figure*}

Our reliable 0.7\,cm images are plotted in Figure \ref{fig1} alongside the 0.7\,cm image from L+98 for comparison. In these data sets, the cycle time was adequate to obtain reasonable phase transfer from the calibrator to the source for the VLA baselines but shorter cycle times were probably required for the other data sets due to poorer weather. We also independently reduced the L+98 0.7\,cm data set and were able to reproduce the notable asymmetry to the East, irrespective of the applied image weighting. Unlike the L+98 image, no large scale asymmetries are present in either of our two 0.7\,cm images as shown in Figure \ref{fig1}, and so the asymmetry reported by L+98 was not a large scale ($\sim 2\,$R$_{\star}$) stable feature.

\subsection{Radio diameters}\label{sec3.2}
\begin{figure}
\includegraphics[trim = 0mm 0mm 0mm 0mm, clip,height=14cm, width=9.0cm]{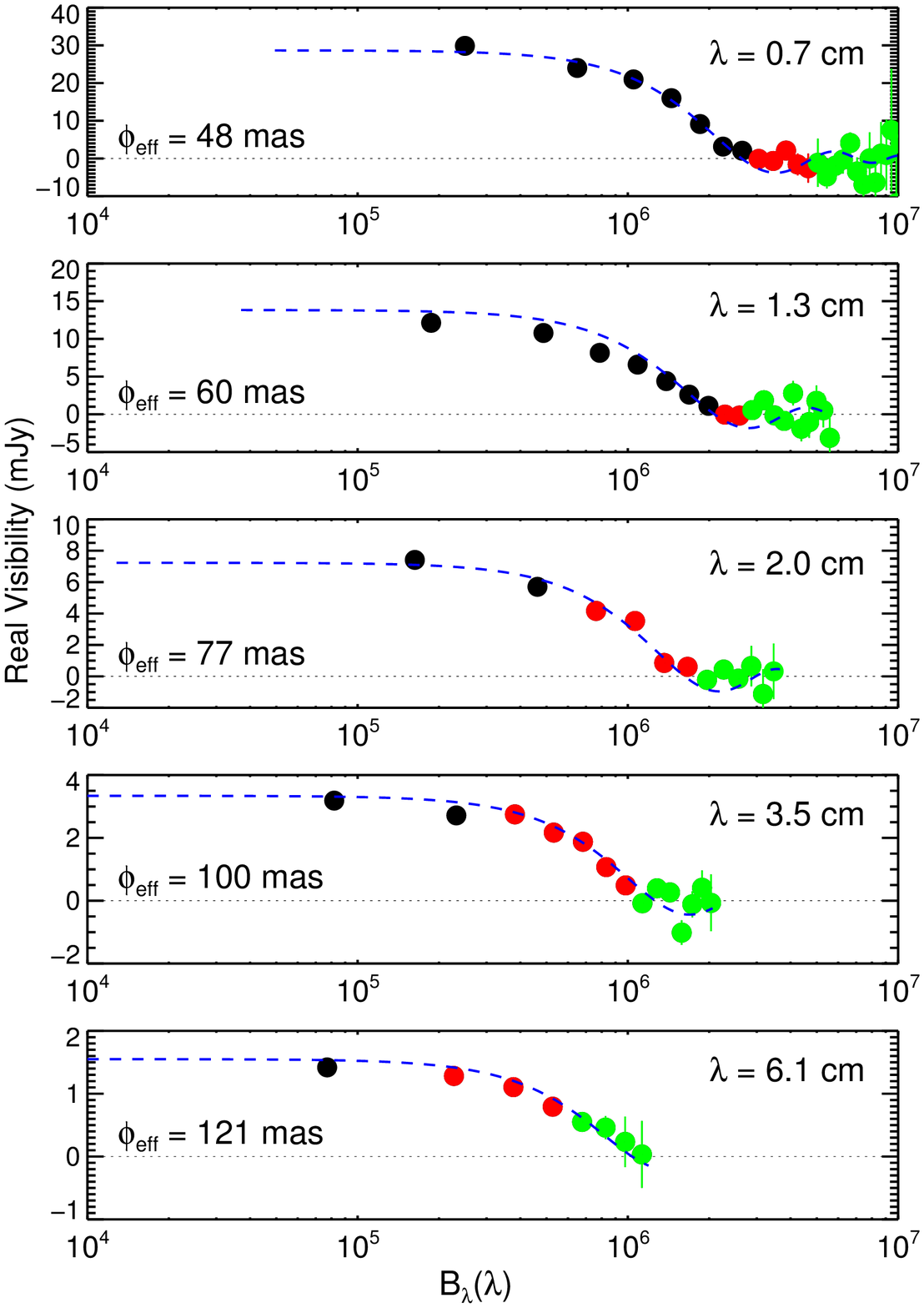}
\caption{Real component of the complex visibility plotted against the projected baseline length in wavelengths, $B_{\lambda}$, for the 2004 epoch at all wavelengths between $0.7-6.1$\,cm. The black data points contain only VLA antenna baselines, the red data points contain both VLA and PT antenna baselines, while the green data points contain only PT antenna baselines. The dashed blue line represents the theoretical visibilities of a uniform intensity circular disk with an \textit{effective angular radius} defined as $\phi _{\rm{eff}}=0.5\sqrt{\theta _{\rm{maj}}\theta _{\rm{min}}}$.}
\label{fig2}
\end{figure}

It can be seen in Figure \ref{fig1} that the size of the source at 0.7\,cm is only a few beam diameters across. The source to beam ratio gets progressively smaller at longer wavelengths which makes fitting models to the images unreliable. In this case it is best to obtain estimates of the source size and flux density by directly fitting to the calibrated complex visibilities. In Figure \ref{fig2} we plot the real component of the complex visibility against the projected baseline length in wavelengths, $B_{\lambda}$, for the 2004 epoch at all wavelengths between $0.7-6.1$\,cm. The data has been azimuthally averaged into uniformly sized bins over large portions of the unevenly sampled $u-v$ plane, where $u$ and $v$ are the spatial coordinates of the projected baseline in wavelengths, and $B_{\lambda} = \sqrt{u^2 + v^2}$. It is clear that the source is resolved at all wavelengths shown, i.e., we have obtained visibilities beyond to the first null. The VLA observations of L+98 only fully resolved Betelgeuse's atmosphere at 0.7\,cm at one epoch, but here we fully resolve the atmosphere at all wavelengths between 0.7 and 6.1\,cm over multiple epochs thanks to the inclusion of the PT antenna. The black data points in Figure \ref{fig2} represent the real visibility measurements from the VLA antenna baselines only, the red data points include both VLA and PT antenna baselines, while the green data points include only PT antenna baselines. This highlights the importance of the inclusion of the PT antenna, especially at the longer wavelengths where it provides many measurements close to the first null, which helps to constrain the size of the atmosphere. The corresponding binned imaginary component of the complex visibilities are not shown in Figure \ref{fig2} but usually had zero values at all baselines, indicative of no large scale asymmetric emission. 

\begin{figure}
\includegraphics[trim = 0mm 0mm 0mm 0mm, clip,scale=0.38,angle=90]{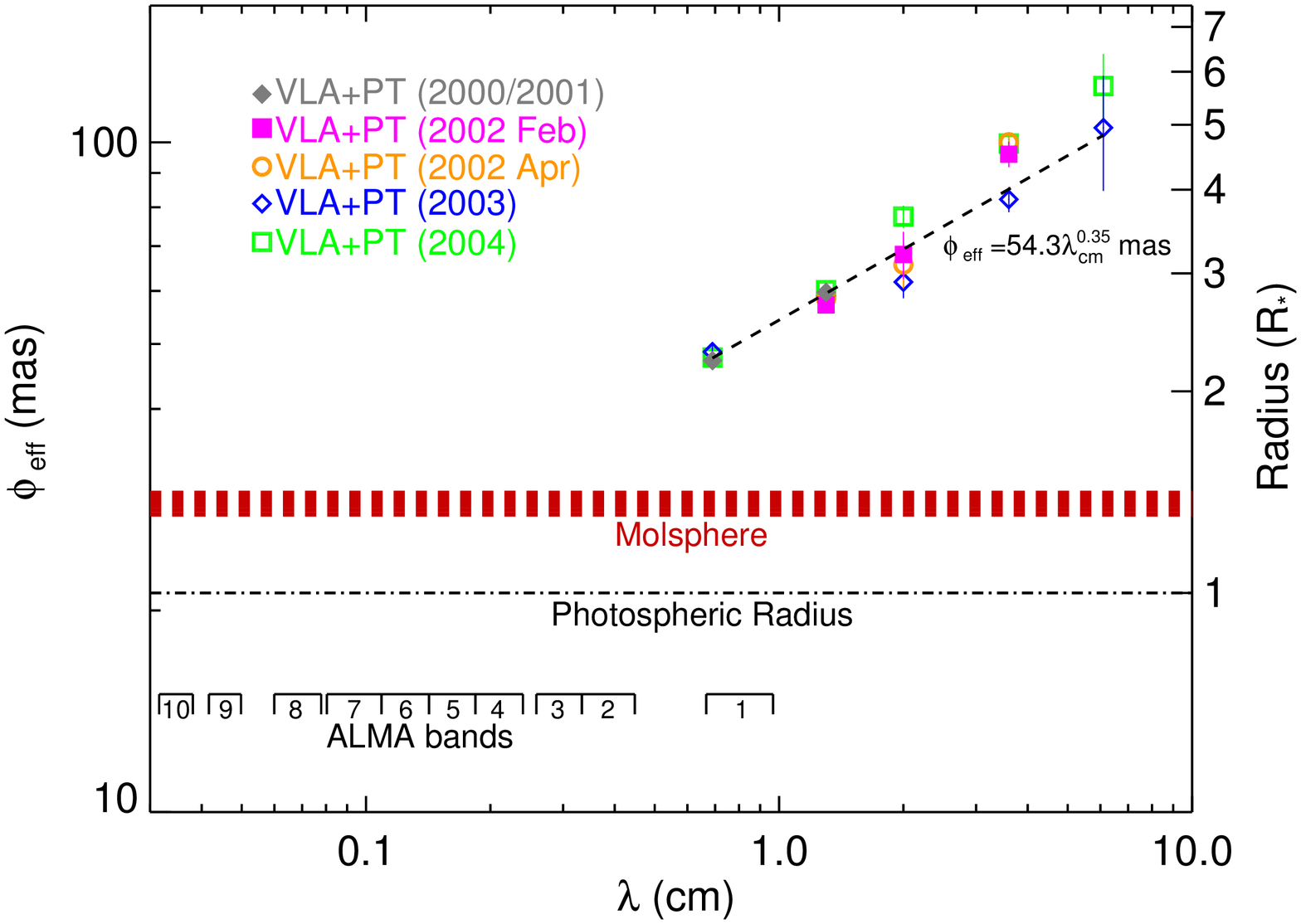}
\caption{The \textit{effective angular radius} of Betelgeuse's extended atmosphere as a function of wavelength. It can be described by the power law, $\phi _{\rm{eff}} = 54.3\lambda ^{0.35} _{\mathrm{cm}}$\,cm, between 0.7 and 6.1\,cm. Multi-wavelength extended ALMA configuration observations will be required to study the morphology of the radio emission within $2\,$R$_{\star}$. The molecular shell (MOLsphere) from \cite{perrin_2007} and the photospheric radius from \cite{ohnaka_2011} are also shown.}
\label{fig2a}
\end{figure}

Uniform intensity disk models were fitted to the complex visibilities as outlined in Section \ref{obs2.2}. The major and minor axes, $\theta _{\rm{maj}}$ and $\theta _{\rm{min}}$, and PAs of the best fits are listed in columns 5, 6, and 7 of Table \ref{tab1}. In Figure \ref{fig2} we include these theoretical visibility models, assuming of a uniform intensity circular disk with an \textit{effective angular radius} defined as $\phi _{\rm{eff}}=0.5\sqrt{\theta _{\rm{maj}}\theta _{\rm{min}}}$. These models provide reasonable though not perfect fits to the azimuthally averaged real visibilities at all epochs. Beyond the first null we can see that there is an even greater departure from the uniform intensity disk model at short wavelengths and this departure is evident at other epochs too. Limb darkening and brightening effects, and a weak unresolved feature located away from the phase center could potentially manifest themselves in the second lobe. Limb brightening could possibly be caused by the temperature rise from the photosphere to the $1.4\,R_{\star}$ temperature peak in the semi-empirical model of \cite{harper_2001}. We find no clear signature in the imaginary component of the complex visibilities or phases at the corresponding baselines for either limb darkening or an additional unresolved weak feature. Indeed, we find that the data provides better fits to the uniform intensity disk model beyond the first null for epochs which have the best $u-v$ sampling and so the consistently poor fit beyond the first null in Figure \ref{fig2} may well be due to the limited $u-v$ sampling.

To further investigate the possible presence of small scale asymmetries in the radio emission morphology, we subtracted the best fit uniform disk model with parameters listed in Table \ref{tab1}, from the corresponding visibilities for each data set. After imaging, most of these uniform disk subtracted data sets consisted of only noise, but a few contained one or two weak but significant (i.e., $3-6\,\sigma _{\mathrm{rms}}$) negative flux density point sources. To derive reliable properties for these point sources and to rule out image deconvolution artefacts, we went back and simultaneously fitted a uniform disk and a delta function to the original calibrated visibilities. For five datasets we found that a uniform disk and a negative flux density delta function offset from disk center produced a slightly better fit than just a single uniform disk. The parameters from these fits are listed in Table \ref{tab2}. The brightness temperatures of the point sources were found to vary between $\sim 300-800\,$K lower than the simultaneously fit uniform disk brightness temperature. Their unresolved nature means that these brightness temperatures are upper limits for the gas temperature, for thermal and optically thick emission. We interpret these negative point sources as pockets in the extended atmosphere where the local gas temperature is lower than the mean gas temperature and highlight the non-uniform nature of Betelgeuse's extended atmosphere.

The centimeter continuum free-free opacity varies as $\kappa _{\lambda} \propto \lambda ^{2.1}$ \citep[e.g.,][]{rybicki_1979} and so our multi-wavelength radio observations probe different layers of Betelgeuse's stellar atmosphere along the line of sight, with the longer wavelengths probing emission further away from the photosphere. \cite{harper_2001} showed that the dominant source of this opacity is from the free-free interactions of free electrons from photoionized metals with neutral hydrogen. We find the source size gets progressively larger with increasing wavelength which implies that the radio surfaces are optically thick, and is in agreement with the findings of L+98. This is clearly illustrated in Figure \ref{fig2a} where the values of the effective angular radius are plotted against wavelength over all epochs between 2000 and 2004. We find that the power law $\phi _{\rm{eff}} = (54.3\pm 0.1)\lambda ^{0.35\pm 0.05} _{\mathrm{cm}}$\,mas can describe the size of the radio emitting region well at all wavelengths between 0.7 and 6.1\,cm as shown by the dashed line in Figure \ref{fig2a}. It is clear from Figure \ref{fig2a} that millimeter and submillimeter spatially resolved observations will be required to study the morphology of the inner atmosphere within 2\,R$_{\star}$. The Atacama Large Millimeter/submillimeter Array (ALMA) will have such capabilities in its more extended array configurations at all wavelengths between Band 2 and Band 10, as shown in Figure \ref{fig2a}. 

We can also investigate if there is any variability in the size of the extended atmosphere over time at the various centimeter wavelengths. At 0.7 and 1.3\,cm there is no significant change in $\phi _{\rm{eff}}$ at all epochs with $\Delta \phi _{\rm{eff}} \lesssim 3\%$, where $\Delta \phi _{\rm{eff}}$ is the maximum deviation about the mean. There is more variability in $\phi _{\rm{eff}}$ at longer wavelengths which sample the more extended atmosphere, with $\Delta \phi _{\rm{eff}} \sim 15\%$ at 2.0 and 3.5\,cm. We only have two measurements at 6.1\,cm but the associated error bars are too large to detect any significant variability in $\phi _{\rm{eff}}$.

\subsection{Flux density variability}
\label{sec3.3}
\begin{figure}
\includegraphics[trim = 0mm 0mm 0mm 5mm, clip,scale=0.39, angle=90]{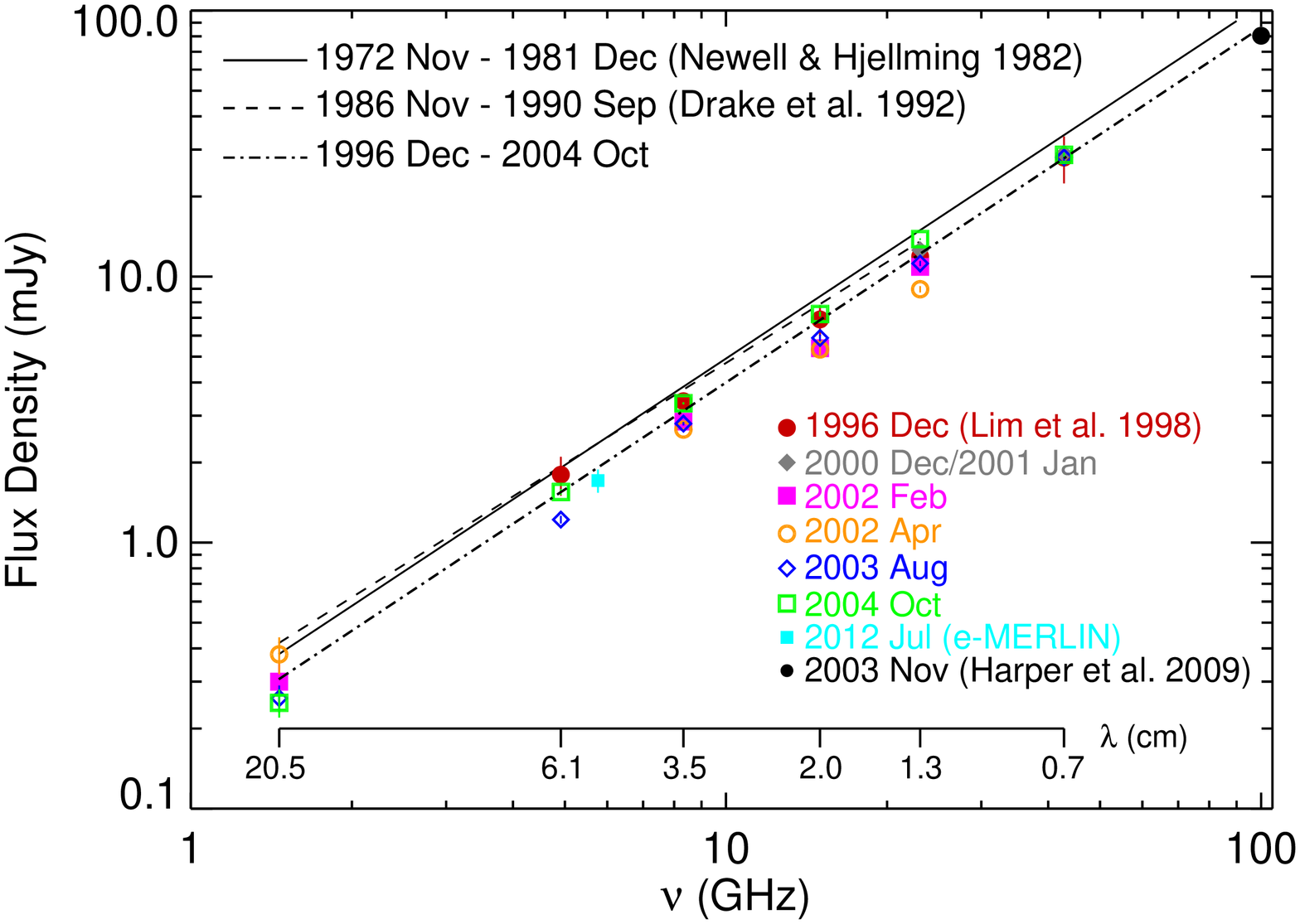}
\caption{Radio spectra for Betelgeuse at VLA wavelengths spanning $1972 - 2012$. The continuous line represents the radio spectrum from \cite{newell_1982} and includes measurements up to 90\,GHz. The dashed line is the almost identical spectrum from  \cite{drake_1992}. The dash-dotted line is the radio spectrum that best fits the more recent data between $1996 - 2004$ and is described by $F_{\nu} = 0.18$\,mJy\,$\nu ^{1.33}$. The flux density from the e-MERLIN 5.2\,cm observation and a 100\,GHz flux density measurement from \cite{harper_2009} are also shown.}
\label{fig3}
\end{figure}

Betelgeuse's atmosphere is optically thick at centimeter wavelengths and so the thermal continuum flux density can be approximated as emanating from an opaque disk with $F_{\nu} \propto T_{\mathrm{gas}}\phi _{\rm{eff}}^2$. The relationship tells us that variations in the radio flux density  are caused by either changes in the mean gas temperature or changes in the size of the extended atmosphere. \cite{newell_1982} found that multi-wavelength radio flux density measurements from the period $1972-1981$ could be described by the radio spectrum $F_{\nu} = (0.23\pm 0.01)\nu ^{1.33\pm 0.01}_{\rm{GHz}}$\,mJy, while \cite{drake_1992} reported an almost identical spectrum for data spanning $1986-1990$.  In Figure \ref{fig3} we plot these radio spectra along with our VLA/PT flux density values and the values from L+98. It can be seen that all the data from 1996 onwards lie below the spectrum of both \cite{newell_1982} and \cite{drake_1992} as initially reported in \cite{harper_2013c} and are better fitted by the spectrum $F_{\nu} = (0.18\pm 0.01)\nu ^{1.33\pm 0.02}_{\rm{GHz}}$\,mJy. Therefore, even though the spectral index, $\alpha$ (where $F_{\nu} \propto \nu ^{\alpha}$), is unchanged, the flux density appears to have reduced by $\sim 20\%$ over the last few decades. 

The previous flux density measurements prior to this work were obtained with single dish instruments and with the VLA in various configurations. \cite{drake_1992} found that their 2\,cm flux densities were systematically lower in the more extended VLA configurations than in the more compact ones, but this was probably a combined result of using long cycle times (up to 17 minutes in duration) and taking the peak of their Gaussian fits for the flux density of a resolved or partially resolved source. Our shorter cycle times (e.g., 3 minutes at 2 cm) along with our method of fitting directly to the visibilities to extract the flux density will avoid these problems and so different VLA configurations cannot account for the recent $\sim 20\%$ reduction in flux density. It is still conceivable that at the very high frequencies, our cycle times might still have been less than the coherence time, and could have resulted in a loss of correlated amplitude. However, at longer wavelengths (i.e., $\geqslant 2.0\,$cm) we still find the flux density to be consistently lower than historical values and so poor coherence cannot account for our lower flux density measurements. Finally, systematic errors are unlikely to account for the $20\%$ reduction in flux density. The absolute flux density uncertainty is only $3-5\%$ at the longer wavelengths, while the main flux density calibrator 3C286, varies by only a few percent every decade \citep{perley_2013}. This would suggest that there has been either a global reduction in the mean gas temperature in Betelgeuse's extended atmosphere or a global reduction in the size of the radio emitting region in the last few decades.

During our multiple epochs of observations, we detect low level flux density variability ($\lesssim \pm 25\%$) at all wavelengths except at 0.7\,cm, which shows no significant variability. This is in excellent agreement with the findings of \cite{drake_1992} who monitored Betelgeuse regularly over a four year period at 2.0, 3.5, and 6.1\,cm and reported similar levels of variability. \cite{drake_1992} also found no correlation between variability at the three wavelengths sampled and noted significant variability on time scales of $\leqslant 1$\,month. Our multi-wavelength data is sparsely sampled over the $\sim 4$\,years but we do find significant variability at 1.3\,cm in just two months. We also find that the flux density increases between our 2003 and 2004 epochs at 2.0, 3.5, and 6.1\,cm and this increase coincides with an increase in the corresponding effective radii. However, a change in flux density does not always coincide with a change in the atmospheres size. For example, the 1.3\,cm flux density variability between the two sampled months of 2002 occurs in spite of the fact that the $\phi _{\rm{eff}}$ remains essentially constant. \cite{bookbinder_1987} used time scale arguments to suggest that this short term variability in the radio flux density could not be caused by large scale processes such as changes in the mass-loss rate or global pulsations because these dynamical time scales are much too long. \cite{bookbinder_1987} suggested this short term variability could be due to the molecular catastrophe scenario of \cite{muchmore_1987} which predicts that global depletions of ions into molecules could rapidly change the radio flux density. This mechanism has been explored by \cite{harper_2001b} who found that either incomplete or large scale formation of CO, SiO, and H$_2$ could be responsible for the observed radio variability triggered by rapid changes in the far ultraviolet radiation field. We propose an alternative mechanism for the flux density variability in Section \ref{disc2}. 

\subsection{Temperature profile of the extended atmosphere}\label{sec3.4}
Spatially resolving the extended atmosphere at multi-wavelengths also allows the brightness temperature to be calculated as a function of depth along the line of sight. If we assume the radio emission is emanating from a uniform intensity disk with major and minor axis, $\theta _{\mathrm{maj}}$ and $\theta _{\mathrm{min}}$, respectively, then the brightness temperature can be written as
\begin{equation}
\label{eq1}
T_{b}=1.96\times 10^{6}\left(\frac{F_{\nu}}{\mathrm{mJy}}\right)\left(\frac{\lambda}{\mathrm{cm}}\right)^2\left(\frac{\theta _{\mathrm{maj}}\theta _{\mathrm{min}}}{\mathrm{mas}^2}\right)^{-1}\,\mathrm{K}\, .
\end{equation}
Since the radio emission from Betelgeuse is thermal and optically thick, the brightness temperature is a measure of the mean gas temperature, $T_{\mathrm{gas}}$, and so multi-wavelength spatially resolved radio observations allow the radial mean gas temperature profile to be constructed.

\begin{figure}
\includegraphics[trim = 0mm 15mm 0mm 0mm, scale=0.38, angle=90]{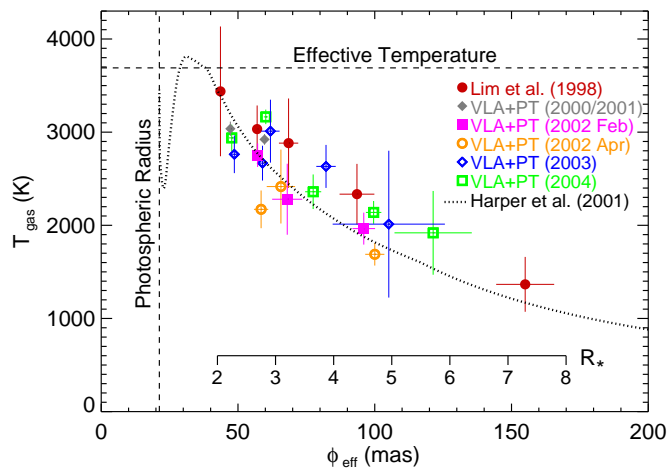}
\caption{The mean gas temperature profile of Betelgeuse's extended atmosphere revealed by multi-wavelength spatially resolved VLA and VLA/PT observations spanning $1996-2004$. The red filled circles are from the work of L+98 while the dotted line is the predicted gas temperature profile from the semi-empirical model of \cite{harper_2001}.	All other data points are from this work. The error bars represent the $\pm 1\sigma _{\mathrm{rms}}$ level, except the 0.7\,cm error bars of L+98 which include a $\pm 20\,\%$ uncertainty in the absolute flux density. The vertical and horizontal dashed lines represent the photospheric radius of $21.24$\,mas and the photospheric effective temperature of $3690$\,K, respectively \citep{ohnaka_2011}}.
\label{fig4}
\end{figure}

In the final column of Table \ref{tab1} we list the brightness temperature at each wavelength and epoch which have been derived using Equation \ref{eq1}. These values are plotted against the effective angular radius, or distance from the photosphere, in Figure \ref{fig4} along with the values from L+98 and the semi-empirical model of \cite{harper_2001}. The gas temperature profiles constructed for all our epochs between 2000 and 2004 are in good agreement with both the values derived from the single epoch of L+98 and the subsequent semi-empirical model of \cite{harper_2001}. Our data implies that the mean gas temperature has a typical value of 3000\,K at 2\,$R_{\star}$, and decreases to  1800\,K at 6\,$R_{\star}$. We do not resolve the atmosphere at 20.5\,cm and so our data does not probe the gas temperature beyond 6\,$R_{\star}$. We also note that shorter wavelength (i.e., $<0.7\,$cm) spatially resolved observations will be required to probe the mean gas temperature of the atmosphere inside $2\,$R$_{\star}$. It is within this region that cool molecular layers of CO, SiO, and H$_2$O exist \citep{tsuji_2000,perrin_2007,ohnaka_2011} and nascent molecule formation may take place, as shown in Figure \ref{fig2a}.

We also detect temporal variability in the gas temperature at certain regions of the atmosphere. At 4.7\,$R_{\star}$ we find the temperature to change by 400\,K over 2.5\,yr while closer in towards the star at 2.8\,$R_{\star}$, we find the temperature to change by almost 600\,K in under two months. These changes are significant, even when we allow for the absolute flux density uncertainties. Nevertheless, most values of the gas temperature are consistent within the random uncertainties implying that the mean gas temperature profile is generally stable with occasional fluctuations on the order of a few hundred Kelvin. Fitting a power-law to all the values displayed in Figure \ref{fig4} yields an empirically derived  temperature profile of the form $T_{\mathrm{gas}}(r)\propto r^{-0.6\pm0.1}$, in good agreement with the power-law fit of  $T_{\mathrm{gas}}(r)\propto r^{-0.7}$ to the semi-empirical temperature profile of \cite{harper_2001}. This temperature profile is less steep than the $r^{-1.65}$ falloff derived for the red giant Arcturus ($\alpha$ Boo: K2\,III) based on model dependent multi-wavelength unresolved centimeter studies \citep{ogorman_2013}. 

\section{Discussion}
\subsection{Where are the e-MERLIN radio hotspots?}\label{disc1}
\begin{figure}
\includegraphics[trim = 10mm 30mm 20mm 30mm, clip,scale=0.55, angle=90]{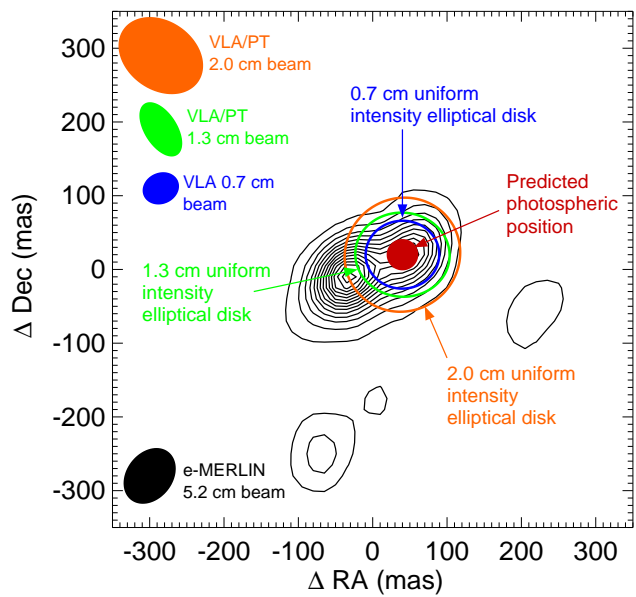}
\caption{Contours represent the e-MERLIN 5.2 cm image of Betelgeuse obtained in 2012 with levels set at $(5,10,15,....,70)\times \sigma _{\mathrm{rms}}$ (where $\sigma _{\mathrm{rms}}=9\,\mu$Jy). The e-MERLIN restoring beam is shown in the bottom left corner, while the restoring beams of our 2004 epoch data at 0.7\,cm, 1.3\,cm, and 2.0\,cm are shown at the top left for comparison. The red filled circle is the scaled photosphere at the 2012 position predicted by the astrometric solution of \cite{harper_2008}. The blue, green, and orange ellipses are the derived sizes of the atmosphere at 0.7\,cm, 1.3\,cm, and 2.0\,cm, respectively, for the 2004 epoch. The combined positional uncertainty is 28\,mas.} 
\label{fig5}
\end{figure}

The multi-epoch spatially resolved VLA/PT data presented in this paper contain no large scale asymmetries in the radio emission morphology, let alone radio hotspots, at any wavelength. In Figure \ref{fig5} we plot the e-MERLIN 5.2\,cm image adapted from R+13 which clearly shows the two radio hotspots, separated by $90\,$mas and having a PA of 110$^{\circ}$ along their line of centers. For comparison, we also include the restoring beams used to create the radio images from our 2004 epoch at 0.7\,cm, 1.3\,cm, and 2.0\,cm. The absence of these radio hotspots in our VLA/PT data at 20.5\,cm, 6.1\,cm, and probably 3.5\,cm could be explained by insufficient spatial resolution. However, our multi-epoch observations at both 0.7\,cm and 1.3\,cm have more than adequate spatial resolution to resolve these features. Moreover, the spatial resolution of our 2.0\,cm VLA/PT images along the direction of the hotspots's line of centres is $\sim 95\,$mas and should also be sufficient to resolve some signatures of the reported hotspots. Since we had the capabilities to spatially resolve the reported radio hotspots at three wavelengths over multiple epochs, we now examine why they have no signature in our VLA/PT data.

The unresolved radio hotspots detected with e-MERLIN at 5.2\,cm (i.e., C band) had flux density values of 0.53 and 0.79\,mJy\,beam$^{-1}$ with a systematic flux density uncertainty of 10\%. In Section \ref{sec3.3} we demonstrated that a spectral index of $\alpha = 1.33$ is appropriate for the total flux density of Betelgeuse at centimeter wavelengths over the three decades prior to 2004. If we assume that the radio hotspots obey this spectral index then we should have detected even the weakest radio hotspot at the 26$\sigma _{\mathrm{rms}}$ level at 1.3\,cm and at the 17$\sigma _{\mathrm{rms}}$ level at 0.7\,cm, in our uniformly weighted images. Assuming the emission to be thermal and taking the extreme case whereby the radio hotspots are optically thin where $\alpha = -0.1$, we would only have just been capable of detecting the weakest radio hotspot at the $3\sigma _{\mathrm{rms}}$ level at 1.3\,cm while it would not have been detectable at 0.7\,cm. However, it would be highly unlikely for the radio hotspots to be optically thin, because in such a case $T_{\mathrm{gas}} >> T_{b}$, and the gas would be fully ionized and thus many orders of magnitude more opaque. A final scenario is the possibility that the emission from one or both radio hotspots is non-thermal and so the spectral index could be very negative and would explain their absence in our high resolution data. However, our 20.5\,cm flux density measurements show no major rise above the expected spectral index value of 1.33 and so we can rule out any significant non-thermal emission. The absence of the radio hotspots in our multi-epoch high spatial resolution data sets cannot be explained by insufficient sensitivity. 

\textit{HST} observations provide us with additional spatial information about Betelgeuse's extended atmosphere. Ultraviolet Faint Object Camera images show a non-uniform brightness distribution and have approximately twice the optical diameter \citep{gilliland_1996}. They therefore probe the same region as our 0.7\,cm radio data presented above. However, no reports from \textit{HST} imaging have revealed features separated by 90\,mas \citep[e.g.,][]{dupree_2013}. While collisionally excited optically thin C~II] 2325$\AA$  emission is observed out to $\pm \sim 4\,$R$_{\star}$, no strong excess emission was detected far from the ultraviolet disk centre \citep{harper_2006}. \textit{HST} spatially scanned spectra reveal radial velocity shifts in ultraviolet spectral features for certain scan angles and not in others. If we interpret the absence of radial velocity shifts to be the scans along the stellar rotation axis, then the PA of the rotational axis is probably $\sim 65^{\circ}$ (see the discussion in \cite{harper_2006} and references therein). This PA does not align with the $110^{\circ}$ PA of the e-MERLIN radio hotspots line of center and so they can not be associated with the rotational pole of the star, where hot spots might be expected due to rotation induced asymmetries \citep[e.g.,][]{uitenbroek_1998}.

The data presented in this paper were obtained between 8 and 12 years prior to when the e-MERLIN radio hotspots were reported. We have also re-reduced the multi-wavelength VLA data from L+98 and can confirm that no radio hotspots were present. The possibility that the extended atmosphere could have undergone such an enormous morphological and thermal reconfiguration in less than a decade after our VLA/PT observations seems remote. Such a change would be in stark contrast to the `radio quiet period' of both our VLA/PT observations and the VLA observations of L+98 which together span a total of $\sim 8$ years and show the atmosphere to be cool, continuously more extended at longer wavelengths, with little or no large scale asymmetries in the morphology. If one (or both) of the radio hotspots were indeed new features, then one would presumably expect a significant increase in the total flux density, but such an increase is in disagreement with the historical flux density measurements described in Section \ref{sec3.3}. 

With such a large amount of multi-epoch multi-wavelength data at our disposal showing no trace of the reported e-MERLIN radio hotspots, one of us (AMSR) has now re-reduced the e-MERLIN data. It has recently been discovered that the position of the Cambridge telescope receiver axis was not entered correctly in the e-MERLIN data during the 2012 Betelgeuse observations \citep{beswick_2015}. This receiver axis offset has now been corrected off-line and preliminary re-imaging at high spatial resolution shows no trace of the previously reported radio hotspots. An examination of the calibration procedure is currently under way to assess the impact of the receiver axis offset correction on the final e-MERLIN data set. Irrespective of this, we can conclude that the radio hotspots as reported in R+13 are most likely interferometric artefacts.

\subsection{Radio flux density versus optical photometry}\label{disc2}
We have also obtained optical V band photometry of Betelgeuse (V = 0.42, B$-$V = $1.85$) as part of our temporal evolution study of its extended atmosphere. This allows us to search for relationships between changes close to the stellar photosphere and changes in the extended atmosphere. The V band observations were taken with 20 and 28\,cm Schmidt-Cassegrain telescopes near Villanova University. An uncooled Optec, Inc. SSP-3 solid state photometer with a silicon red-sensitive PIN-photodiode detector was used to record raw photometer counts through a V band filter centered on 550\,nm. The filter had a similar spectral response to the standard Johnson V band filter. The photometry itself was conducted differentially with respect to the K0\,IIIb comparison star HD\,37160 (V = 4.09, B$-$V = 0.95). The comparison star was further observed differentially with respect to the B2\,III check star HD 35468 (V = 1.64, B$-$V = $-0.22$) to check the non-variability of the comparison star. Apart from random magnitude scatter, there was no variation between the comparison and check stars.

\begin{figure*}
\includegraphics[trim = 0mm 5mm 0mm 0mm, clip,width=13cm, height=18cm, angle=90]{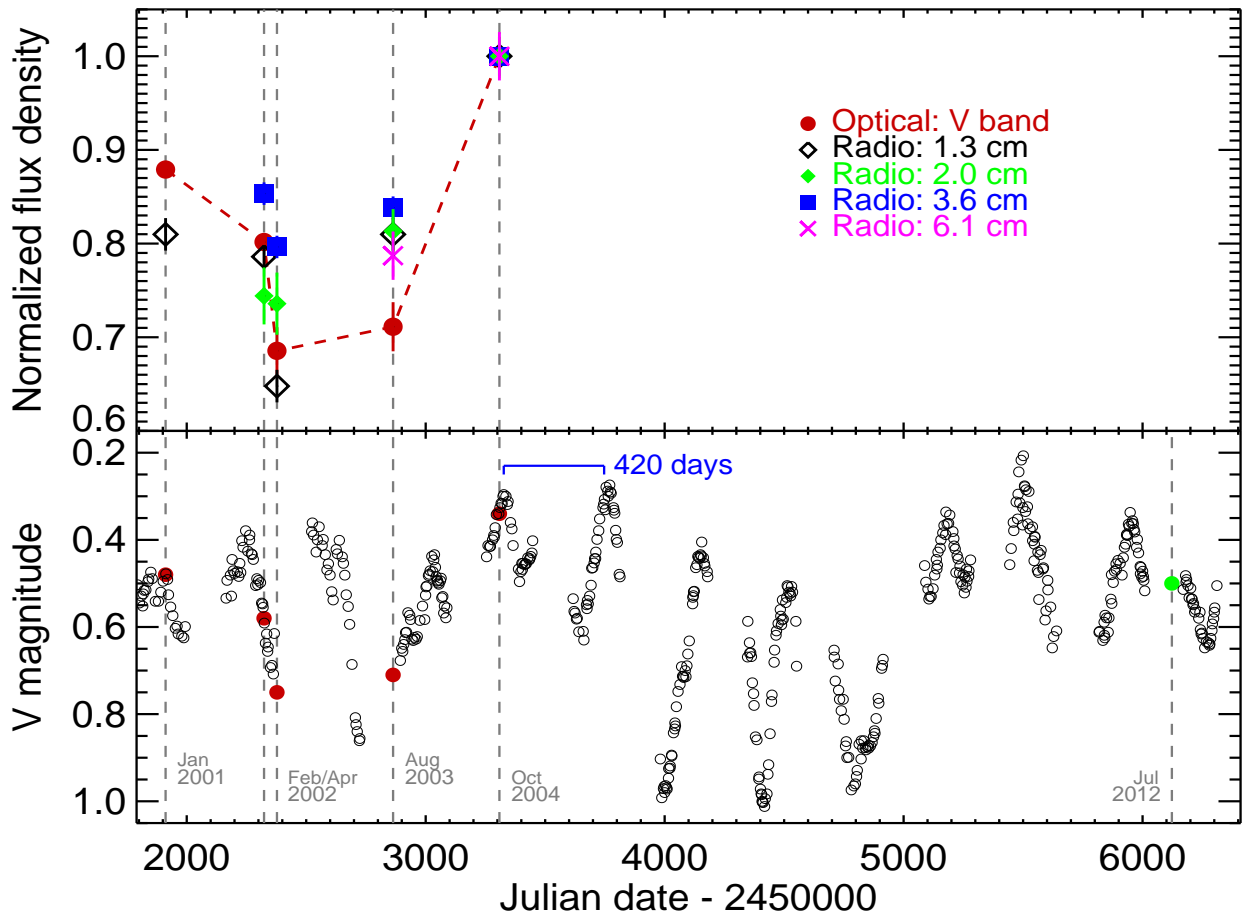}
\caption{\textit{Top panel}: The variation of V band optical photometry with our radio flux density values for all wavelengths between 1.3 and 6.1\,cm. The optical photometry has units of flux density and has been normalized to the October 21-30 2004 value. The radio flux densities at each wavelength have also been normalized to their corresponding October 21-30 2004 value. \textit{Bottom panel}: The original V band optical photometry in units of V magnitude. The red and green filled circles are the V band photometric data at the corresponding dates of the radio observations. An example of the well known $\sim 420$ day period is included which is believed to originate from stellar pulsations. The dashed lines in both panels should guide the eye.} 
\label{fig6}
\end{figure*}

In the bottom panel of Figure \ref{fig6} we plot our V band photometry of Betelgeuse spanning from September 2000 to January 2013. The periodic gaps in the photometry are due to the fact that optical observations of Betelgeuse can only be acquired at Villanova from September through to early April. The optical photometry has good overlap with our 5 epochs of VLA/PT radio data and has simultaneous coverage for 3 epochs, with the final two epochs only 10 days (April 2002) and thirty days (August 2003) away from the nearest photometric values. Estimates of the photometric values for the epochs with non-simultaneous coverage were obtained using the SLICK method within the PERANSO software package \citep{vanmunster_2014}. The SLICK method subtracts the peak from the time series data and then carries out a Fourier transform of the residual spectrum iteratively until a good fit is found. The red filled circles plotted in the bottom panel of Figure \ref{fig6} are the V magnitude values for the dates corresponding to our VLA/PT observations, while the single green filled circle corresponds to the date of the e-MERLIN observations. We simply note that there was nothing striking with the optical photometry around the time of the e-MERLIN observations and focus the remainder of this discussion on comparing the VLA/PT data with the V band data.

Betelgeuse is a semiregular variable with both photometric and radial velocity studies revealing two prominent periods; a short primary period of $\sim 400$ days \citep[e.g.,][]{smith_1989} and a longer secondary period lasting $\sim 2000$ days \citep[e.g.,][]{sanford_1933, spencer_jones_1928}. The long secondary period may be related to the convective turnover time of hypothetical giant convection cells on the stellar surface \citep{stothers_1971, stothers_2010}, although we see no clear evidence for it in our 12 years of photometry. The shorter primary period has been observed photometrically \citep{dupree_1987} and in radial velocity data \citep{smith_1989, dupree_1990} and is consistently evident over the 12 years of our photometry. This primary period displays a large range of peak-to-peak amplitude variation over 12 years with $0.3 \leq \Delta V \leq 0.7$, which is a similar range of values to what were reported during the mid-1980s, but larger than the variation reported during the early 1990s \citep{krisciunas_1992}. We can see from the bottom panel of Figure \ref{fig6} that our VLA/PT radio data (i.e., red filled circles) spans about 4 different primary periods, at various phases for each of these periods. To compare the photometry and the radio data, we have converted the V magnitudes to flux densities and normalized both the radio and V band flux densities to their values for the 2004 VLA/PT epoch. We note that we exclude our three 0.7\,cm measurements from this analysis as they show no significant variability over their three epochs. The results are shown in the top panel of Figure \ref{fig6}. It can be seen that there is a similar behaviour between the variability of the optical and the radio flux densities. 1.3\,cm measurements are available for 5 epochs, and an increase or decrease in the radio flux density is accompanied by a change of optical flux density in the same direction between all pairs of epochs. The other radio wavelengths are measured at fewer epochs but show similar trends, although in some cases with low significance, e.g., between the closely-spaced second and third epochs. This shows that there is an irregular but distinct correlation on timescales of a few months.

To understand why the optical and radio flux density might exhibit similar variability trends, we discuss the origin of the $\sim 400$ day primary period. \cite{dupree_1987} discovered a 420 periodic photospheric (B band; $\lambda=453$\,nm) and chromospheric (both in the ultraviolet continuum at $\lambda = 300\,$nm and in the \ion{Mg}{II} \textit{h} and \textit{k} line emission cores) modulation of the flux density. They argued that this periodic modulation could not be caused by rotation due to the required unrealistic rapid rotation rates, nor could it be due to a few hot convection cells as the emergence of such cells would not be expected to be periodic. Periodic photospheric pulsations were deemed to be the most likely candidate and a time lag of $\sim 70$ days in chromospheric variation after photospheric variations suggested the presence of travelling waves in the atmosphere. Radial velocity observations from \cite{smith_1989} confirmed a primary period of $400\pm 20$ days and again associated its origin to stellar pulsations. V band photometry covers a large number of TiO band heads, whose formation is very temperature sensitive. The V band intensity change can be interpreted to result from changes in the optical flux caused by temperature changes in the photosphere \citep{morgan_1997}. However, the strengths of optical TiO band heads may also be affected by temperature effects from the chromosphere \citep{lobel_2000}. Assuming the longer wavelength radio emission originates in the more extended atmosphere, one would expect a delay of $\sim 3-6$ years for a disturbance to propagate out from the photosphere to these radio formation regions, assuming a velocity of 10\,km\,s$^{-1}$ \citep[e.g.,][]{ogorman_2012}. Clearly, this is too long a timescale in comparison to the similar timescales between both the optical and the radio flux density variability that is evident in Figure \ref{fig6}. 

Pulsation induced radiation field changes must be invoked to explain these relative quick changes. In such a scenario, pulsation induced shocks heat the inner layers of the extended atmosphere, resulting in a local increase in the electron density. Recombination (with timescales on the order of a few days e.g., \citealt{harper_2001b}) increases the global radiation field causing an almost instantaneous increase in the photoionization of metals in the more extended atmosphere. This process can account for large-scale and fast changes in the radio fluxes through changes in the radio opacity (i..e, $\kappa _{\lambda} \propto n_{\mathrm{e}}n_{\mathrm{ion}}$). We can see in Figure \ref{fig6} that our 1.3\,cm flux density measurements track changes in the optical photometry well. We find no variation in the size of the extended atmosphere at this wavelength and so the flux density variability at this wavelength must be due to gas temperature changes. \cite{harper_2001} predict that most of the 1.3\,cm emission is formed within $\sim 1.8\,R_{\star}$ and so the shocks might damp within this region. This is in good agreement with lower mass long period variables whose pulsation induced shocks damp within 2\,$R_{\star}$ \citep{reid_1997}. The pulsation induced changes in the radio opacity at longer wavelengths might then be the cause of the variations in the extended atmosphere's effective angular radius, which is obvious in Figure \ref{fig2a}. It remains unclear why the 0.7\,cm flux density remains almost constant, but it may well be a result of the relatively large absolute flux density uncertainty and the possible loss of coherence at this short wavelength. The absence of correlated variations in the 7\,mm flux densities may also be a result of the higher densities and heat capacity of the emitting region, which leads to smaller temperature changes as compared to longer wavelengths.

\section{Conclusions}
We have used the VLA along with the Pie Town VLBA antenna to study the extended atmosphere of Betelgeuse at 0.7, 1.3, 2.0, 3.5, 6.1, and 20.5\,cm, over multiple epochs between 2000 and 2004. Combining the Pie Town antenna with the most extended configuration of the VLA has allowed us to obtain visibilities beyond the first null at all wavelengths between 0.7 and 6.1\,cm inclusive, implying that we fully resolve the extended atmosphere at these wavelengths. We find that the extended atmosphere deviates from circular symmetry at all wavelengths and that there may exist small pockets of gas significantly cooler than the mean global temperature of the extended atmosphere. We also find that the radio flux density from Betelgeuse has decreased by about 20\% in comparison to measurements from the 1970s and 1980s, despite the spectral index remaining almost constant at $\alpha = 1.33$. This implies that there has been a global reduction in the gas temperature and/or size of the extended atmosphere. 

We find that both the size and temperature profile of Betelgeuse's extended atmosphere as a function of wavelength is in good agreement with the findings of L+98. At 0.7\,cm the extended atmosphere is over twice the size of the optical photosphere, while at 6.1\,cm it increases to about six times the size. We confirm that Betelgeuse's extended atmosphere has a low mean gas temperature over multiple years. The mean gas temperature has a typical value of 3000\,K at 2\,R$_{\star}$ and decreases to 1800\,K at 6\,R$_{\star}$. The temperature profile can roughly be described by the power law $T_{\mathrm{gas}}(r)\propto r^{-0.6}$ although temporal variability of a few 100\,K are evident at some epochs. 

The e-MERLIN 5.2\,cm unresolved radio hotspots detected in 2012 have no signature in our VLA/PT data spanning from 2000 to 2004 even though we had sufficient spatial resolution and sensitivity at short wavelengths to detect them. Radio continuum emission is biased towards regions that are hot and ionized, the two most often going together, so one would expect signatures at multiple wavelengths. The combined constraints provided by the multiple epochs of spatially resolved multi-wavelength VLA/PT data has encouraged us to re-examine the e-MERLIN data. From this we can conclude that the e-MERLIN radio hotspots described in R+13 are most likely interferometric artefacts.

We report a similar behaviour between the variability of V band optical photometry and the radio flux density at certain wavelengths. The optical variability is known to be induced by stellar pulsations, and so the radio flux density variability is likely a manifestation of stellar pulsation induced shocks. These shocks might damp at distances within 1.8\,$R_{\star}$ and cause variations in the structure of Betelgeuse's extended atmosphere through pulsation induced radiation field changes. The similarities between the variation of the optical and the radio flux densities suggest that stellar pulsations could play an important role in exciting the extended atmospheres of red supergiants. 

\begin{acknowledgements}
The data presented in this paper were obtained with the  Very Large Array (VLA) which is an instrument of the National Radio Astronomy Observatory (NRAO). The NRAO is a facility of the National Science Foundation operated under cooperative agreement by Associated Universities, Inc. EOG and GMH acknowledge support from Science Foundation Ireland grant SFI11/RFP.1/AST/3064 and a grant from Trinity College Dublin. EOG and WV acknowledge support from ERC consolidator grant 614264.
\end{acknowledgements}

\bibliographystyle{aa}
\bibliography{references}

\begin{appendix}
\section{Deviations from a uniform disk} \label{app1}
In Table \ref{tab2} we present the PT data sets where a uniform intensity disk plus one or more delta functions provided a better fit over a single uniform disk model. All of these six data sets contained negative point sources in the residual images after the best fit uniform disks from Table \ref{tab1} were subtracted from their visibilities.
\label{app1}

\begin{table*}
\caption{PT observations best fit by a uniform intensity disk plus one or more delta functions.}
\label{tab2}
\centering
\begin{tabular}{c c c c c c c c c}
\hline\hline
Date						& $\lambda$			    & Shape            & $F_{\nu}$      		& $\theta _{\mathrm{maj}}$  &  $\theta _{\mathrm{min}}$	& PA  & Offset$^{\dagger}$  & $T_{\mathrm{b}}$
 \\
     			& (cm)                        		&     	& (mJy)                    	& (mas)   		    & (mas)   		    & ($^{\circ}$)   		    & (mas, mas)   		& (K)    \\
\hline
\rule{-2.6pt}{2.5ex} 2004 Oct 21,30  
		 				 & 1.3 & Disk & $16.16\pm 0.37$   & $148\pm 4$& $ 106\pm 4$ & $88\pm 4$& $\dots$ & $3410\pm 175$\\
		 				 &    & Delta & $-1.21\pm 0.21$   & $\dots$& $\dots$ & $\dots$& $54\pm 4, -1\pm 5$ & $2585\pm 355$\\
		 				 &    & Delta & $-0.99\pm 0.20$   & $\dots$& $\dots$ & $\dots$& $-53\pm 5, -3\pm 7$ & $2735 \pm 345$\\
		 				 & 2.0& Disk & $7.96\pm 0.26$     & $173\pm 8$ & $ 145\pm 11$ & $126\pm 18$& $\dots$ & $2490\pm 235$\\
		 				 &    & Delta & $-0.66\pm 0.20$   & $\dots$& $\dots$ & $\dots$& $48\pm 13, -32\pm 20$ & $2165\pm 375$\\
\hline
\rule{-2.6pt}{2.5ex}  2003 Aug 10,12& 2.0		& Disk& $6.83\pm 0.34$	& $146\pm 13$& $116\pm 19$& $145\pm 21$& $\dots$ & $3010\pm 340$\\
									 & 			& Delta& $-0.85\pm 0.27$	& $\dots$& $\dots$& $\dots$& $35\pm 15, -25\pm 24$ & $2395 \pm 330$\\
\hline
\rule{-2.6pt}{2.5ex}  2002 Feb 17,18	& 3.5		& Disk &	$3.30\pm 0.12$ &$225\pm 11$ & $179\pm 22 $& $ 130\pm 13$ & $\dots$ & $1965\pm 270$\\
									& 			& Delta &	$-0.44\pm 0.10$&$\dots$ & $\dots$& $ \dots$ & $-67\pm 18, 3\pm 30$ & $1695 \pm 360$\\
\hline
\rule{-2.6pt}{2.5ex}  2000 Dec 23 & 0.7		& Disk & $30.00\pm 0.40$	& $100\pm 2$ &$90\pm 2$ &$160\pm 7$ & $\dots$ & $3200\pm 105$\\
					 & 			& Delta& $-0.94\pm 0.25$	&$\dots$ & $\dots$ & $\dots$ & $28\pm 3, -27\pm 4$ & $2490 \pm 320$\\
\hline
\end{tabular}
\tablefoot{$^{\dagger}$ The positional offset of the delta function from the center of the uniform disk. The first term is the offset in RA with a positive value denoting an offset to the east while the second term is the offset in Dec with a positive value denoting an offset to the north.}
\end{table*}

\end{appendix}

\end{document}